\documentclass[prl,twocolumn,superscriptaddress,letterpaper]{revtex4}
\usepackage{graphicx,amssymb,bm,natbib,amsfonts,amsmath}

\begin{document}

\title{Charge-Carrier Screening in Single-Layer Graphene}

\author{David A. Siegel}
\affiliation{Department of Physics, University of California,
Berkeley, CA 94720, USA}
\affiliation{Materials Sciences Division,
Lawrence Berkeley National Laboratory, Berkeley, CA 94720, USA}

\author{William Regan}
\affiliation{Department of Physics, University of California,
Berkeley, CA 94720, USA}
\affiliation{Materials Sciences Division,
Lawrence Berkeley National Laboratory, Berkeley, CA 94720, USA}

\author{Alexei V. Fedorov}
\affiliation{Advanced Light Source, Lawrence Berkeley National Laboratory, Berkeley, CA 94720, USA}

\author{A. Zettl}
\affiliation{Department of Physics, University of California,
Berkeley, CA 94720, USA}
\affiliation{Materials Sciences Division,
Lawrence Berkeley National Laboratory, Berkeley, CA 94720, USA}

\author{Alessandra Lanzara}
\affiliation{Department of Physics, University of California,
Berkeley, CA 94720, USA}
\affiliation{Materials Sciences Division,
Lawrence Berkeley National Laboratory, Berkeley, CA 94720, USA}

\begin{abstract}
The effect of charge carrier screening on the transport properties of a neutral graphene sheet is studied by directly probing its electronic structure. We find that the Fermi velocity, Dirac point velocity and overall distortion of the Dirac cone are renormalized due to the screening of electron-electron interaction in an unusual way  We also observe an increase of the electron mean free path due to the screening of charged impurities.  These observations help us to understand the basis for the transport properties of graphene, as well as the fundamental physics of these interesting electron-electron interactions at the Dirac point crossing.
\end{abstract} 

\maketitle

Unlike normal metals where charge carriers and impurities are highly screened by the Fermi sea, the introduction of charges to a neutral graphene sheet has several competing effects on its transport properties, due to the screening of the electron-electron interaction and the screening of long-range impurities (such as charges or vacancies).  
While the former is expected to renormalize the Fermi velocity and the Dirac point velocity~\cite{DasSarma2007}, the latter could lead to a decrease of the quasiparticle scattering rate or an increase in the electron mean free path.~\cite{Monteverde2010,CastroNeto2009}  
These effects are important for those applications, such as spintronics, where the electron mean free path is a more relevant quantity than the conductivity.\cite{Tombros2007}  
Although the effect of electronic screening has been extensively studied in the past~\cite{Katsnelson2006,Stauber2007,Katsnelson2008,deJuan2010, Ostrovsky2006, Hwang2008, CastroNeto2009, Favaro2006, DasSarma2011, Borghi2009, Brey2009, DasSarma2007,Adam2007}, revealing unusual behavior upon the introduction of charge carriers on the graphene sheet \cite{Chen2008}, most of these works have focused on the renormalization of the Fermi level \cite{Li2008,Elias2011}, and cannot address the important question of how the various electronic effects can renormalize the Dirac cone or otherwise contribute to conduction.  

Angle-resolved photoemission spectroscopy (ARPES) is an ideal tool to probe the electronic properties of graphene~\cite{Hufner2003, Bostwick2007, Zhou2007, McChesney2010, SiegelCopper, SiegelPNAS, Hwang2012}.
In many of these works however, the starting graphene layer is highly doped, and as a consequence the effect of charge-carriers on the electronic screening is more difficult to discern.  Similarly, in the case of undoped graphene, the focus has been on how the dielectric screening of the substrate has an impact on the electronic dispersions~\cite{SiegelPNAS,Hwang2012}.  
Therefore, the question of how charge carrier screening affects the Dirac cone dispersion and how it differs from dielectric screening remains an open questions.  

Here we demonstrate the effects of charge-carrier screening on a graphene sheet: with the progressive deposition of small quantities of potassium, we observe a singularity in the Fermi velocity and Dirac point velocity; an overall renormalization of the valence band; a decrease in the quasiparticle scattering rate; and qualitatively different behaviors from the case of dielectric screening.  These results demonstrate the many ways in which charged impurities can have an impact on the transport properties of graphene.

The h-BN samples were prepared by CVD growth on a Cu film, followed by transfer to the h-BN substrate and hydrogen annealling.  The graphene was placed on mechanically exfoliated flakes of h-BN, many layers high (opaque to visible light), which were in turn supported on a doped Si wafer with native oxide.  The sample preparation was nearly identical to that described in past references (the samples in our experiment were not patterned).~\cite{Decker2011,Gannett2011,Dean2010}
Our ARPES investigation was performed at beamline 12.0.1.1 at the Advanced Light Source, at a pressure better than 3$\times$10$^{-11}$ torr, with a sample temperature of 15 K, and photon energy 50eV.  The sample was annealed to 700$^\circ$C in UHV prior to measurement.  The sample was electron-doped in situ by potassium deposition with an SAES Getters alkali metal dispenser at 15K, under which conditions the potassium atoms sit above the graphene surface in a disordered arrangement.~\cite{Bennich, McChesney2010}  The in situ deposition allowed us to study the same position on the sample as potassium was progressively added.

Much attention has recently been focused on the properties of hexagonal boron nitride (h-BN) as a substrate for graphene electronics~\cite{Decker2011,Dean2010,Xue2011,Gannett2011}.  Graphene/h-BN has significantly improved transport properties and fewer charged impurities than previously studied graphene/SiO$_2$ systems~\cite{DasSarma2011a}.  ARPES spectra of graphene/h-BN are shown in Figure 1.  Following the maximum intensity, one can clearly observe nearly linear energy spectra, characteristic of Dirac electrons\cite{Geim2007}.  As potassium is added to the sample, the Dirac point appears and moves to higher binding energy (indicated by the black arrows in panels d-h), and the charge density of the sample increases.

As potassium is added the spectral widths do not increase significantly, which is surprising since impurities often broaden ARPES spectra.  In Figure 2 we examine this effect, showing the widths of momentum distribution curves (MDCs, intensity profiles as a function of momentum) for different dopings.  Panel (a) shows that the spectral widths vary almost linearly as a function of binding energy, but have an overall offset: the sample with higher charge density (larger k$_\mathrm{F}$) has smaller (sharper) MDC widths.

To quantify this further, in panel (b) we plot the widths of MDCs at the Fermi level, as a function of the Fermi momentum $k_\mathrm{F}$.  The width of an MDC at the Fermi level is proportional to the quasiparticle scattering rate or inverse mean free path of the photohole\cite{CastroNeto2009,Siegel2008}, and $k_\mathrm{F}$ is proportional to the square root of the charge density.  We observe that as the carrier concentration increases in absolute magnitude, the MDC widths decrease.  This behavior differs from short-range impurity scattering, where scattering rates are expected to increase with charge density.\cite{Peres2006}  The decrease in the MDC widths also cannot be due to other many-body effects, which generally have vanishing contributions to the spectral widths at the Fermi level.\cite{Park2009}  Therefore the screening of long-range impurity scattering is the only remaining explanation for the sharpening of spectral features: when the increase in charge density improves the screening of long-range impurity potentials, the quasiparticle scattering rate is expected to decrease with increasing doping.\cite{deJuan2010}

\begin{figure}
\includegraphics[width=8.5 cm] {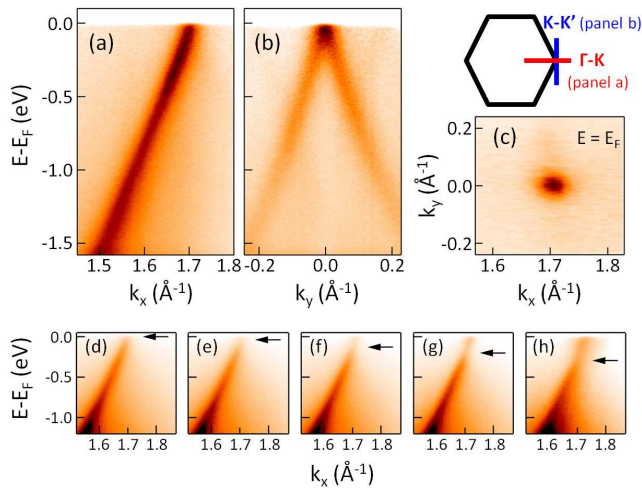}
\caption{ARPES spectra of graphene on h-BN.  (a) ARPES spectrum of single-layer graphene along the $\Gamma$-K direction.  This is the direction along which all of our data is analyzed in subsequent figures.  (b) ARPES spectrum along the K-K' direction (perpendicular to $\Gamma$-K) shows that the Dirac point is at the Fermi level.  The spectra in panels (a)-(b) have been normalized by the area under the MDCs.  (c) The pointlike Fermi surface of graphene. (d-h) Doping dependence along the $\Gamma$-K direction, with Fermi k-vectors corresponding to k$_\mathrm{F}$ = -0.0035, 0.0037, 0.0126,0.0206, 0.0282 $\mathrm{\AA}^{-1}$, respectively.}
\end{figure}

\begin{figure} \includegraphics[width=8.5cm]{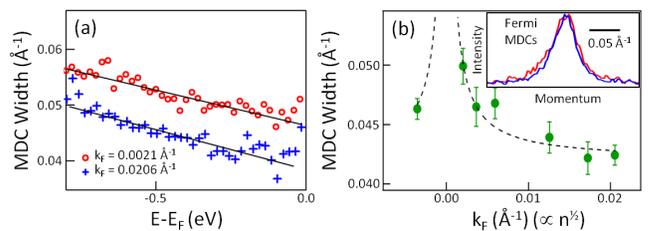}
\caption{The screening of charged impurities can lower the quasiparticle scattering rate.  (a) The imaginary self-energy (Im$\Sigma$), proportional to the width of ARPES spectra and inversely proportional to the quantum lifetime of the photohole, is shown for two different dopings ($n$ is proportional to $k_\mathrm{F}^2$).  In addition to the roughly linear binding energy dependence for each spectrum is an overall offset between them: the spectra with added potassium have \textit{sharper} spectral features than the as-grown sample, a naively counterintuitive finding.  (b) A comparison of the MDC widths at the Fermi level shows that increasing the charge density sharpens spectral features by screening the interaction between quasiparticles and impurities.  The dashed line is a 1/k$_\mathrm{F}$ fit to the data, which allows us to extract an impurity density of $n_\mathrm{imp} \sim 1.6 \times 10^{11} \mathrm{cm}^{-2}$.  The inset shows the raw MDCs at the Fermi level, confirming that the blue curve is sharper than the red.}\end{figure}

Figure 2b can be compared to theoretical calculations of the scattering rate from long-range impurities\cite{deJuan2010,Stauber2007}.  The dashed line shows a fit to a 1/$k_\mathrm{F}$ behavior, where the constant of proportionality gives the impurity density:
\begin{equation} \mathrm{Im} \Sigma = \alpha^2 n_\mathrm{imp} v_\mathrm{F} \pi I(2 \alpha) / k_\mathrm{F} + Const.\label{eq1}\end{equation}
Here Im$\Sigma$ is the imaginary self energy, $\alpha$ is the effective fine-structure constant of graphene, $n_\mathrm{imp}$ is the impurity density, $v_\mathrm{F}$ is the Fermi velocity, $I(2 \alpha)$ is a dimensionless constant (we use $I(2 \alpha) \approx 0.22$)\cite{Hwang2008}, and $Const$ is an overall offset.  Using $\alpha = 0.78$ (discussed below) and $v_\mathrm{F} = 0.85 \times 10^6 \mathrm{m/s}$ (the bare LDA velocity), we find $n_\mathrm{imp} = (1.94\pm0.37) \times 10^{11} \mathrm{cm}^{-2}$, which is typical for graphene/h-BN~\cite{DasSarma2011a}, smaller than the impurity density of graphene on SiO$_2$ (typically $\geq 10^{12} \mathrm{cm}^{-2}$)~\cite{Tan2007}, and an order of magnitude smaller than the potassium density of the highest doping in figure 1 ($1.3 \times 10^{12}$, assuming $\sim$ 1 electron donated per potassium atom~\cite{KReference}).

\begin{figure*} \includegraphics[width=18cm]{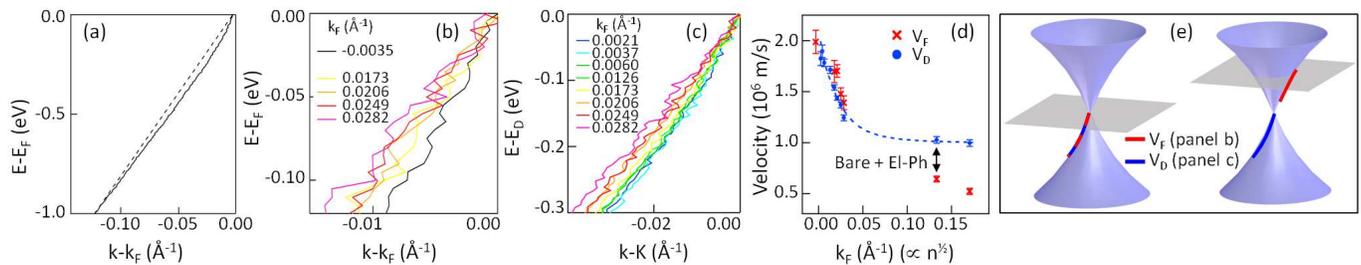}
\caption{Velocity renormalization by charge-carrier screening.  (a) An electronic energy-momentum dispersion extracted from the peak positions of the MDCs from undoped graphene/h-BN is shown as the solid black curve.  The curvature is due to the long-range electron-electron interaction~\cite{SiegelPNAS}, and can be compared to the (arbitrary) straight dashed line in the same figure.  (b) Extracted dispersions show the Fermi velocity as a function of doping.  For ease of viewing, the momenta of these dispersions have been aligned so that the Fermi k-vectors coincide (the x-axis is given as k-k$_\mathrm{F}$).  (c) Extracted dispersions show the Dirac point velocity as a function of doping.  For ease of viewing, the energies of these dispersions have been aligned so that the Dirac points coincide (the y-axis is given as E-E$_\mathrm{D}$).  (d) The Fermi velocities and Dirac point velocities are given as a function of doping ($n$ is proportional to k$_\mathrm{F}^2$).  Both the Dirac point and Fermi level show velocity enhancements as the Dirac point approaches the Fermi energy.  The dashed line is a logarithmic fit to the Dirac point velocities; see text for further details. (e) These two Dirac cones give a schematic of where the data is taken in panels (b) and (c).  Panel (b) shows dispersions near the Fermi level (marked in red in panel e), which change position with respect to the Dirac point as the Fermi level is adjusted.  Panel (c) shows dispersions near the Dirac point, which remain at the same place on the Dirac cone even as the Fermi energy is altered.}\end{figure*}

\begin{figure} \includegraphics[width=8cm]{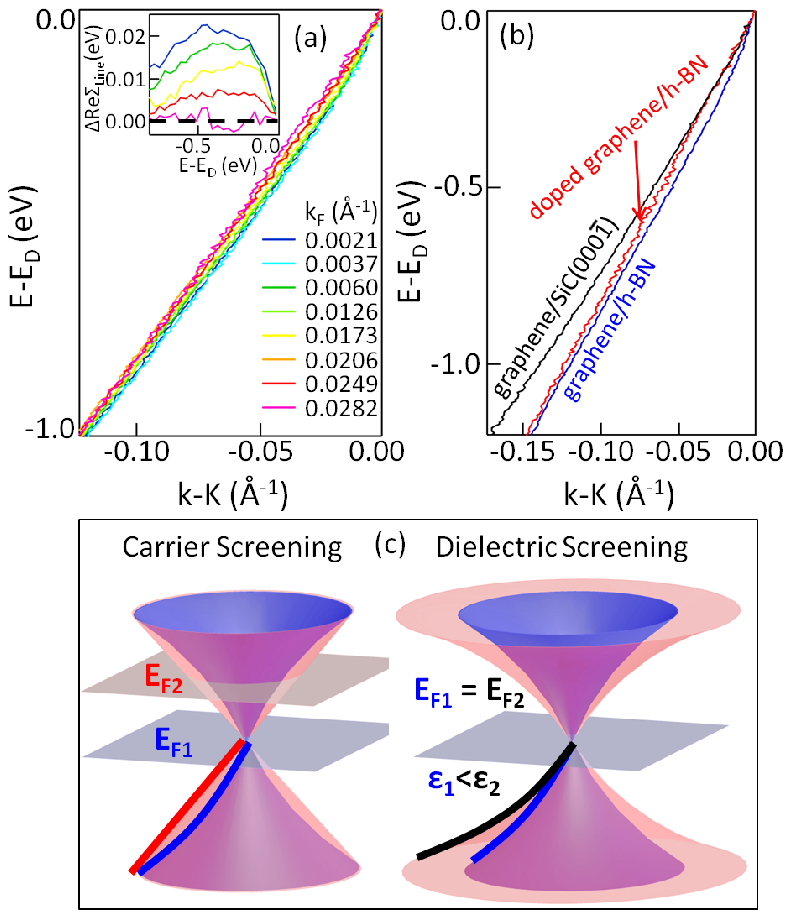}
\caption{Dirac cone renormalization by charge-carrier screening.  (a) Electronic dispersions near the Dirac point show higher binding energy behavior than figure 3c.  The inset shows the difference between the experimental bands and an arbitrary straight line.  Panel (b) illustrates the difference between charge carrier screening and dielectric screening.  Three dispersions are shown: (i) as-grown graphene/h-BN; (ii) as-grown graphene/SiC($000\overline{1}$); and (iii) doped graphene/h-BN.  The doping for (iii) was chosen so that the band velocity near the Dirac point would match that of (ii).  The cartoons in panels (c) and (d) illustrate the renormalization effect on the Dirac cones for charge-carrier screening and dielectric screening, respectively.  In panel (c), the renormalization is primarily restricted to low momenta.  In panel (d), the renormalization extends to all momenta within our range of measurement, and becoming larger in magnitude at higher momenta.}\end{figure}

To further investigate the effect of long-range screening, in Figure 3 we report the doping dependence of the graphene bandstructure.  The electronic energy-momentum dispersion of graphene can be obtained by fitting Lorentzian lineshapes to the MDCs and extracting peak positions as a function of energy.  For as-grown graphene/h-BN, the extracted dispersion has been displayed as the solid black line in figure 3a, illustrating the logarithmic velocity enhancement in the vicinity of the Dirac point.


As the charge density increases, the Fermi velocity decreases.  The band dispersions near the Fermi level are plotted for several dopings in panel 3b.  The Fermi velocity, proportional to the slope of the ARPES dispersions, decreases by a factor of 4, with a maximum of $2.0\times10^6$ m/s, as the Fermi k-vector increases by two orders of magnitude.  These results have been plotted as the red open circles in panel 3d and confirm the results of past experiments.~\cite{Li2008,Elias2011}

However, there are several effects that cooperate to reduce the Fermi velocity as a function of doping.  For instance, the bare band of graphene along the $\Gamma$-K direction is known to decrease in velocity away from the Dirac point~\cite{Park2008}, with a Van Hove singularity at higher dopings.~\cite{McChesney2010}  Given that the band dispersion in panel 3a demonstrates a logarithmic divergence near the Dirac point, even a rigid shift of the Fermi energy would lead to a logarithmic Fermi velocity dependence.~\cite{SiegelPNAS}  Finally, the electron-phonon interaction is also known to renormalize the Dirac cone as a function of doping.~\cite{Calandra2007}

Therefore, in order to separate the electron-electron renormalization from other many-body and bare-band effects, in figure 3c we show the electronic dispersions in the vicinity of the Dirac point as a function of doping, and the velocities have been plotted in panel 3d.~\cite{Chae2012}  The Fermi velocity and Dirac point velocity roughly coincide for low dopings.  However, as the charge density increases the effects of the bare band velocity and electron-phonon renormalization become significant, causing the Fermi velocity to be only half as large as the Dirac point velocity.  Overall, for the purposes of understanding the electron-electron interaction, the Dirac point velocity may be the most fundamental quantity.~\cite{DasSarma2007}

The Dirac point velocity can be fitted with a logarithmic dependence\cite{DasSarma2007, Elias2011}:
\begin{equation}v_\mathrm{D}^* = v_\mathrm{D} \frac{r_\mathrm{s}^0}{4 (1 + a k_\mathrm{F}^2)} \ln(\frac{k_\mathrm{c}}{k_\mathrm{F}}) + Const.\label{eq6}\end{equation}
where $r_\mathrm{s} \equiv e^2 / v \varepsilon$ gives the value of the dielectric constant, and the constant $a$ introduces the same fit parameter as Ref~\cite{Elias2011}, allowing $\varepsilon$ to effectively increase with charge density.  From the fit we obtain $a = 720 \mathrm{\AA}^2$, and $v_\mathrm{D} r_\mathrm{s}^0 / 4 = (0.168 \pm 0.014) \times 10^6 \mathrm{m/s}$.  Using the bare LDA value of $v_\mathrm{D} = 0.85 \times 10^6 \mathrm{m/s}$, we obtain $\alpha = 0.78$ or $\varepsilon^0 = 3.3$.  This value of $\varepsilon$ compares well with 
the reported logarithmic fit to the binding energy dependence in undoped graphene/h-BN~\cite{Hwang2012} where 
$\varepsilon^0 = 4.22$; and with the expected dielectric screening, being $\varepsilon^0 = (\varepsilon_\mathrm{h-BN}+1)/2 = 4.02$~\cite{Geick1966}.

In figure 4, we observe the differences between two separate types of electronic screening effects: screening by the graphene charge carriers, which has been modified in this study by changing the number of charge carriers through potassium deposition; and screening by the dielectric environment, which can be modified by changing the dielectric substrate~\cite{Hwang2012}.  In both cases the Fermi level and Dirac point velocities are modified, varying linearly with $\varepsilon$, and logarithmically with $k_\mathrm{F}$ or doping.  On the other hand, for charge carrier screening the inverse screening length q$_\mathrm{s}$ varies linearly with k$_\mathrm{F}$ and with 1/$\varepsilon$, given by~\cite{CastroNeto2009}:
\begin{equation}\label{eq:screeninglength}
q_\mathrm{s} = 4 \alpha k_\mathrm{F}.
\end{equation}
One might therefore expect differences between these screening effects to be observed at high values of momentum, where k$\gg$q$_\mathrm{s}$.

So in figure 4a we compare the doping-dependence of the graphene valence band dispersions over a larger range of energy and momentum than figure 3c.  At lower values of momentum (near the Dirac point), the band velocities decrease as a function of doping; but at higher momenta this trend begins to reverse, with increasing velocity as a function of doping near 0.1 $\AA^{-1}$.  In contrast, figure 4b (and Ref.~\cite{Hwang2012}) shows that when the band dispersions of graphene on different dielectric substrates are compared, increasing the dielectric constant leads to uniformly smaller band velocities at any given value of momentum or energy.

To make this comparison more straightforward, figure 4b shows three graphene dispersions: (i) as-grown graphene/h-BN; (ii) as-grown graphene/SiC($000\overline{1}$); and (iii) doped graphene/h-BN.  (i) and (ii) have different substrates and therefore different dielectric environments; (i) and (iii) have the same substrate but different charge carrier concentrations; and (ii) and (iii) have different dielectric screening and charge carrier concentrations.  While (ii) and (iii) have similar band velocities in the vicinity of the Dirac point, overlapping for small values of k, these dispersions diverge for larger values of k.  For k$\gg$k$_\mathrm{F}$ the extent of the renormalization (or magnitude of the self-energy) is found to be strongly dependent on the dielectric constant $\varepsilon$, but weakly dependent on the screening by charge carriers or k$_\mathrm{F}$.  This confirms that the electron-electron interaction is indeed a long-range interaction, with a variable length scale due to the concentration of free charges in graphene.  The ways in which charge carrier screening and dielectric screening modify the Dirac cone are illustrated in figures 4c and 4d, respectively.

In conclusion, we have demonstrated some of the detailed ways in which the addition of charge carriers to a graphene sheet can have an effect on transport properties and the renormalization of the Dirac cone.  The electron-electron and electron-impurity interactions are found to be long-range interactions, and in both cases the addition of charge carriers is shown to decrease the length scale and strength of the interaction.  The increase in charge density is also shown to renormalize the Dirac cone in a distinct manner from dielectric screening.  These results illustrate the differences between charge-carrier screening and dielectric screening in graphene, illuminating the transport behavior of graphene while demonstrating the interesting differences between the electronic interactions of graphene and those of ordinary metals.

\section{Acknowledgments}
DAS would like to thank Chris Jozwiak for help with manuscript preparation, and Marco Polini and Allan MacDonald for useful discussions.

This work was supported by the Director, Office of Science, Office of Basic Energy Sciences, Materials Sciences and Engineering Division, of the U.S. Department of Energy under Contract No. DE-AC02-05CH11231.

Correspondence and requests for materials should be addressed to ALanzara@lbl.gov.

\end{document}